\documentclass[11 pt]{article}
\usepackage{graphicx}
\usepackage{amsmath}
\usepackage{amssymb}
\usepackage{times}
\usepackage{subfigure}
\usepackage[sort,square]{natbib}
\usepackage{setspace}
\usepackage{float}

\floatstyle{ruled}
\newfloat{model}{thp}{lop}
\floatname{model}{Model Excerpt}

\newtheorem{theorem}{Theorem}

\newtheorem{lemma}[theorem]{Lemma}
\newtheorem{definition}[theorem]{Definition}

\def\thisPaperTitle{ Knowledge Flow Analysis for Security Protocols}
\title{\thisPaperTitle}

\author{
Emina Torlak, Marten van Dijk, Blaise Gassend, Daniel Jackson, and Srinivas Devadas \\
\{emina, marten, gassend, dnj, devadas\}@mit.edu}

\begin{document}
\maketitle


\begin{abstract}
Knowledge flow analysis offers a simple and flexible way to find flaws in security protocols. A protocol is described by a collection of rules constraining the propagation of knowledge amongst principals. Because this characterization corresponds closely to informal descriptions of protocols, it allows a succinct and natural formalization; because it abstracts away message ordering, and handles communications between principals and applications of cryptographic primitives uniformly, it is readily represented in a standard logic. A generic framework in the Alloy modelling language is presented, and instantiated for two standard protocols, and a new key management scheme.

\end{abstract}

\section{Introduction}

One area of major successes for formal methods has been the verification of security protocols. A number of specialized tools have been developed in the last decade that have exposed subtle flaws in existing protocols (see, e.g. \cite{Lowe97, CJM00}). For the most part, however, these tools have been used by the researchers that developed them, and less attention has been paid to usability issues.

This paper presents a new approach to formulating and checking cryptographic protocols. It does not enable any new form of analysis. Instead, it makes verification more accessible to the designers of protocols. Its key contribution is a new characterization of these protocols that is both closer to how designers conceive them, and amenable to a more direct encoding in standard first-order logic. This more direct encoding allows existing tools to be applied as black boxes without modification; it requires no tweaking of parameters or issuing of special directives by the user. Moreover, because the semantic gap between informal descriptions of protocols and their formalization is smaller, there are fewer opportunities for errors to creep in.

In this paper, the Alloy modeling language is used to record the details of the protocol and its security goals, and the Alloy Analyzer is used to find flaws. The approach, however, requires no special features of Alloy or its analysis, and could be applied in the context of any formal method based on first-order logic. Its simplicity suggests that it may be useful in teaching; indeed, using the approach, we have explained cryptographic protocols to undergraduates who have had only a few weeks of experience in formal methods.

Our approach, which we call {\em knowledge flow analysis}, gives a uniform framework for
expressing the actions of principals, assumptions on
intruders, and properties of cryptographic primitives. The dynamic behaviour of the protocol is described by an initial state of knowledge, and a collection of rules that dictate how knowledge may flow amongst principals. A state is given by a relation mapping principals to the values they know; the allowable knowledge flows can thus be succinctly described as a standard transition relation on knowledge states, written as a constraint.

This simple setup allows us to model a range of intruder capabilities and to detect replay, parallel session, type flaw, and binding attacks. We have applied it to both symmetric and public-key cryptography under the Dolev-Yao \cite{DY83} approach. The modeling framework itself is more general, however, and can be extended to include the properties of cryptographic primitives \cite{ComonShmatikov03IntruderDeductionsConstraintSolvingInsecurity, ChevalierETAL03NPDecisionProcedureProtocolInsecurityXOR, Shmatikov04ProtocolsProductsModularExponentiation, MillenShmatikov04SymbolicProtocolAnalysisAbelianGroup} and an unbounded number of sessions with bounded messages \cite{ChevalierETAL03ExtendingDolevYaoIntruderAnalyzing}.

This approach grew out of an effort to check a new cryptographic scheme \cite{Gassend03PhysicalRandomFunctions, GassendETAL02ControlledPhysicalRandomFunctions}. Knowledge flow analysis described here was the final result of a series of incremental attempts at formalizing and checking the protocol using the Alloy language and tool. This process helped crystallize our intuitions, and drew out a number of important assumptions. The final analysis, although only performed over a finite domain, actually establishes the correctness of the protocol for unbounded instantiations because of a special property of this protocol. The Alloy models developed for this case study were generalized into a simple framework that was subsequently applied to some standard protocols, such as Needham-Schroeder \cite{NS78} and Otway-Rees \cite{Otway-Rees87}.

The contributions of this paper are:

\newcounter{icount}
\begin{list}{\arabic{icount}.}{\usecounter{icount}} 
\item the knowledge flow formalism, which characterizes the dynamic behaviour of a cryptographic protocol in terms of the increasing knowledge of the principals, avoiding the need to impose an explicit ordering on messages;
\item a realization in the Alloy modelling language as a generic framework with a library of primitives that can be easily instantiated for a variety of protocols;
\item soundness and completeness results that guarantee that (1) any counterexample generated by the analyzer to a security theorem is legitimate, and not an artifact of the modelling framework, formalism or analysis; and (2) that if a counterexample exists involving any number of message exchanges and any number of steps, it will be found, so long as the number of parallel sessions is within a prescribed bound;
\item case study applications of the approach to two well-known protocols, one of which (Needham-Schroeder) is explained in detail, and to a new key management scheme based on controlled physical random functions \cite{Gassend03PhysicalRandomFunctions, GassendETAL02ControlledPhysicalRandomFunctions}.
\end{list}

Section 2 explains the key intuitions underlying the approach, using Needham Schroeder as an example. Section 3 shows the complete formalization of this example, including the statement of the security goal, and a discussion of the counterexample corresponding to the well-known attack. Section 4 gives a mathematical summary of the approach without reference to any particular modeling language that might serve as a basis for implementations in other tools, and which makes precise the assumptions underlying the model. The paper closes with an evaluation and a discussion of related work.
 
\section{Knowledge Flow Basics}
\begin{figure*}
\centering
\includegraphics[scale=0.8]{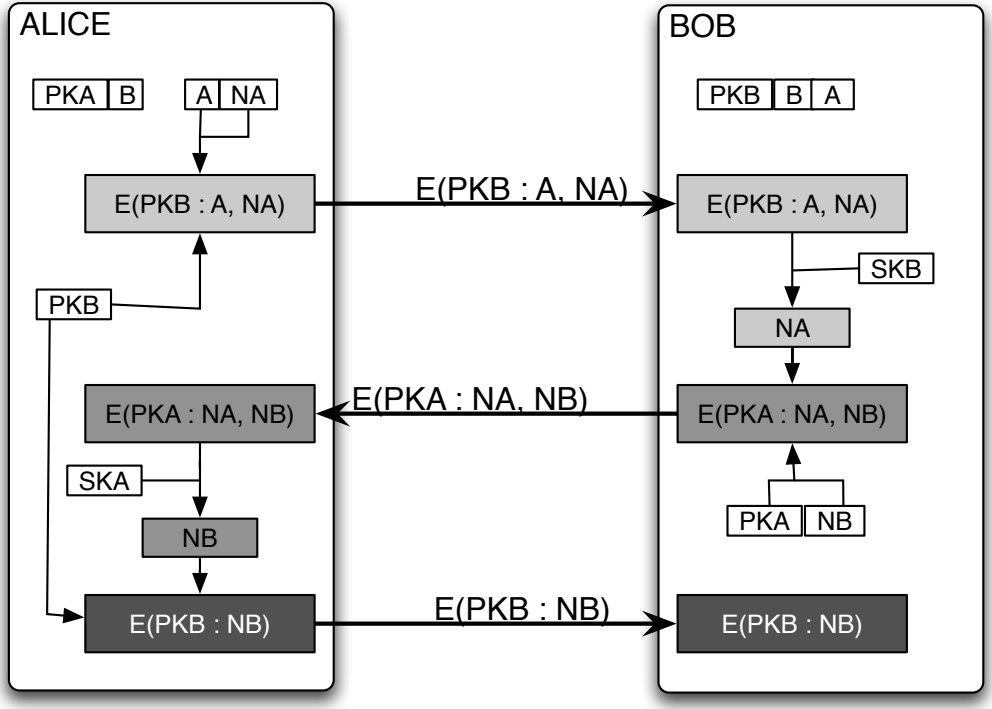}
\caption{Knowledge Flow in Needham-Schroeder Protocol}
\label{ns-diagram}
\end{figure*}

The key idea behind knowledge flow analysis is the observation that, at the most basic level, the purpose of a security protocol is to distribute knowledge among its legitimate participants.  A protocol is flawed if it allows an intruder to learn a value that is intended to remain strictly within the legitimate principals' pool of knowledge.   To gain more intuition about knowledge flows in security applications, consider the Needham-Schroeder Public Key Protocol \cite{NS78} shown in Figure \ref{ns-diagram}. 

We have two principals, Alice and Bob, each of whom has an initial pool of knowledge represented with white boxes.  Alice's initial knowledge, for example, consists of her own public/private key pair $PK(A)$/$SK(A)$, identity $A$, nonce $N_A$, and Bob's public key $PK(B)$ and identity $B$.  The purpose of the protocol is to distribute the nonces between Alice and Bob in such a way that the following conditions hold at the end:  (1) Alice and Bob both know $N_A$ and $N_B$, and (2) no other principal knows the nonces.     

To initiate the protocol, Alice first expands her pool of knowledge to include $E_{PK(B)}(A, N_A)$, an encryption of her identity and nonce with Bob's public key.  She then sends the cipher to Bob who decrypts it using his private key, $SK(B)$.  At the end of the first step of the protocol, each principal's knowledge has increased to include the values in light gray boxes.  Bob performs the second step of the protocol by adding $E_{PK(A)}(N_A, N_B)$ to his current knowledge and sending the cipher to Alice.  She uses her private key to decrypt Bob's message and extract $N_B$.  By using $N_B$ and $PK(B)$, Alice can set up an authenticated and private channel with Bob as is done during the final step of the protocol in which Alice creates $E_{PK(B)}(N_B)$ and forwards it to Bob.  Both Alice and Bob now know the two nonces and share all other knowledge except their secret keys.

Following the flow of knowledge in the Needham-Schroeder protocol provides a crucial insight underlying our analysis method.  Namely, a principal can learn a value in one of three ways;  he can
\begin{list}{$\cdot$}{\setlength{\itemsep}{-3pt} }
\item \textit{draw} the value at the start,  
\item \textit{compute} it using his current knowledge, or 
\item \textit{learn} it by communication.  
\end{list}
Our analysis treats the latter two ways of obtaining knowledge as equivalent.  Specifically, we can think of Alice's computing $E_{PK(B)}(A, N_A)$ as her learning it from a principal called \textit{Encryptor} whose initial pool of values includes all possible ciphers:  Alice sends the tuple $(PK(B), (A, N_A))$ to \textit{Encryptor} who responds by sending back the encryption of $(A, N_A)$ with $PK(B)$.  

Treating cryptographic primitives as principals allows us to consider the total pool of knowledge to be \textit{fixed}.  That is, the set of all values before and after the execution of a security protocol is the same; the only difference is the distribution of those values among the principals.  Since we assume that principals never forget values, the set of principals who know a value at the end of a protocol session subsumes the set of principals who drew the value at the beginning.      

The goal of analyzing knowledge flows in a protocol is to verify that particular values never leak out of  the honest participants' pool of knowledge.  In other words, \textit{we are interested in analyzing the flow of knowledge from an intruder's perspective}.  This observation allows us to make sound simplifying assumptions that drastically reduce the effort needed to formalize a protocol in terms of knowledge flows:  
\begin{list}{$\cdot$}{\setlength{\itemsep}{1pt} }
\item We need not encode the flows of knowledge among the honest principals, such as  the flow which allows Alice to learn $E_{PK(A)}(N_A, N_B)$ from $Encryptor$.  Rather, we may assume that each honest principal draws all values in the total knowledge pool and specify protocols solely in terms of the intruders' knowledge flows (sections \ref{kf-basic-defs} and \ref{initial-knowledge}).  
\item We may model all adversaries, including the untrusted public network, with a single opponent whom we call $Oscar$.  The soundness of this approach is formally proved in section \ref{adversary}.  Intuitively, the approach makes sense if we note that the potential adversaries will be most effective when they collaborate and share knowledge among themselves.  Hence, we can replace the (collaboration of) adversaries with a single principal who possesses all their knowledge, without excluding any intrusion scenarios.           
\end{list}

In our example, the flow of knowledge from the intruder's perspective starts with the protocol initialization message $E_{PK(B)}(A, N_A)$, since Oscar needs no prior knowledge to learn the first cipher that Alice sends to Bob.  In general, because Oscar includes the untrusted public network, he learns the first message of the protocol for free, regardless of who its intended recipient and sender are: 
\begin{equation}\forall_{p\in \{a,b\}, p'\in \{a,b\}\cup O} \ [\emptyset \rightarrow E_{PK(p')}(I(p),N(\epsilon,I(p)))].\label{ns1}
\end{equation}
The variables $a$ and $b$ denote the honest principals (Alice and Bob), and the set $O$ stands for Oscar.  The notation $N(\epsilon,I(p))$ represents the nonce that the nonce primitive $N$ generated for the principal identified by $I(p)$ using the random value $\epsilon$ as the seed.  For example, Alice's identity is $I(a)=A$ and Alice's nonce is
$N(\epsilon,I(a))=N_A$. The empty set means that Oscar does not need prior knowledge to learn $E_{PK(p')}(I(p),N(\epsilon,I(p)))$.

Once his pool of knowledge includes $E_{PK(B)}(A, N_A)$, Oscar learns the corresponding response, $E_{PK(A)}(N_A, N_B)$.  More generally\footnote{We use the parameter $v$ in $c$ instead of $N(\epsilon,I(p))$ because $p'$, the recipient of $c$, cannot conclusively determine that $v$ is, in fact, the nonce $N(\epsilon,I(p))$.},
\begin{eqnarray}
\forall_{p'\in \{a,b\}, p\in \{a,b\}\cup O, v\in V} [c \rightarrow E_{PK(p)}(v,N(c,I(p')))] \label{ns2}  \\
\mbox{where } c = E_{PK(p')}(I(p),v). \nonumber
\end{eqnarray}
The variable $V$ denotes the set of all values, or the fixed pool of knowledge.  Note that our formalization constrains the seed of Bob's nonce to be Alice's initialization message.  This is needed to establish that Bob's nonce was generated in the context of the protocol session started by Alice with $E_{PK(B)}(A, N_A)$.  The resulting correspondence between the nonces prevents our analysis from sounding false alarms when Oscar legitimately obtains two nonces from Alice and Bob by running a valid protocol session with each.

Oscar learns the final message, $E_{PK(B)}(N_B)$, as a consequence of knowing $E_{PK(A)}(N_A, N_B)$.  Formally,
\begin{eqnarray}
 \forall_{p\in \{a,b\}, p'\in \{a,b\}\cup O, v\in V} \nonumber \\ 
 \left [\{E_{PK(p)}(N(\epsilon,I(p)),v))\}  \rightarrow E_{PK(p')}(v) \right].\label{ns3}
\end{eqnarray}

\section{Example}\label{example}

The Needham-Schroeder protocol is vulnerable to a parallel session attack discovered by Gavin Lowe \cite{Lowe96}.  This section presents a knowledge flow analysis of the protocol that reproduces Lowe's results, and gives a flavor of the expressiveness and simplicity of our method.  We have encoded the knowledge flows in the Alloy modelling language \cite{Jackson02AlloyTOSEM} and used the Alloy Analyzer \cite{Jackson00AutomatingFOL} to find the attack.  However, the modelling pattern presented here is applicable to any first-order logic with relations and transitive closure.  

\subsection{Encoding Basic Entities and Relations}

The basic components of a knowledge flow model are the sets $Principal$ and $Value$, and the relations $draws$, $learns$, and $knows$ (Model Excerpt \ref{basicdeclarations}).  

\newcounter{M1count}
\begin{model}[ht]
\centering
\begin{list}{\scriptsize{\arabic{M1count}}}{\usecounter{M1count} \setlength{\itemsep}{-4pt} 
\setlength{\leftmargin}{13pt} \setlength{\rightmargin}{-6pt}  } 
\tt \frenchspacing
\item module kf/basicdeclarations
\item
\item {\bf abstract sig} Value \{\} \label{Value}
\item {\bf sig} CompositeValue {\bf extends} Value \{\} \label{CompositeValue}
\item {\bf sig} AtomicValue {\bf extends} Value \{\} \label{AtomicValue}
\item 
\item {\bf abstract sig} Principal \{ \label{Principal}
\item \  draws: {\bf set} Value, \label{draws}
\item \  owns: {\bf set} draws \label{owns}
\item \}\{ {\bf no} owns \& (Principal - this).@owns \} \label{owns-constraint}
\item 
\item {\bf sig} HonestUser {\bf extends} Principal \{ \label{HonestUser}
\item \}\{ draws = Value \} \label{draws-constraint} 
\item 
\item {\bf one sig} Oscar {\bf extends} Principal \{ \label{Oscar}
\item \  knows: {\bf set} Value, \label{knows}  
\item \  learns: knows->knows \label{learns}
\item \}\{ {\bf no} \verb+^+learns \& iden \} \label{learns-constraint}
\item
\item {\bf pred} InitialKnowledge() \{ \label{InitialKnowledge}
\item \ no CompositeValue \& Oscar.draws \}
\item
\item {\bf pred} FinalKnowledge() \{ \label{FinalKnowledge}
\item \ {\bf all} v: Value | 
\item \ \ v {\bf in} (Oscar.draws).*(Oscar.learns) {\bf iff} 
\item \ \ v {\bf in} Oscar.knows \}
\end{list}

\caption{Generic Model of Principals and Values}
\label{basicdeclarations}
\end{model}

The set $Principal$ includes all principals in a protocol -- the legitimate protocol participants, represented by the subset $HonestUser$, and the intruders, represented by $Oscar$.  The set $Value$ models the fixed pool of knowledge on which a protocol operates.  We distinguish between $AtomicValue$s, which are uninterpreted, and $CompositeValue$s, which may consist of other values and are learned by communicating with cryptographic primitives.  In the example from Figure \ref{ns-diagram}, Alice and Bob are members of $HonestUser$; $Value$ consists of the union of values enclosed in the boxes `Alice' and `Bob'; the identifiers $A$ and $B$ are $AtomicValue$s, and the ciphers are $CompositeValue$s.             

The relation $draws$ (line \ref{draws}) maps each principal to the set of values known by that principal at the beginning of the protocol.  For example, both Alice and Bob draw Alice's identity $A$ at the start of the protocol session shown in Figure \ref{ns-diagram}.  The declaration of $owns$ (line \ref{owns}) together with the constraint on line \ref{owns-constraint} relate a principal to the set of drawn values which uniquely identify him.  Bob, for instance, $owns$ his identity, $B$, even though both he and Alice draw it.   

The field $knows$ (line \ref{knows}) defines the set of all values that Oscar can learn by using the knowledge flows available to him; this includes the knowledge obtainable from both the protocol rules and the cryptographic primitives.  The acyclic relation $learns$ (lines \ref{learns}-\ref{learns-constraint}) encodes the partial ordering on Oscar's maximal knowledge, enforced by the flows from which the knowledge was acquired.  For example, the protocol rule \ref{ns2} specifies that Oscar learns $E_{PK(A)}(N_A, N_B)$ from $E_{PK(B)}(A, N_A)$.  Hence, {\tt Oscar.knows} contains both ciphers and {\tt Oscar.learns} includes the mapping \begin{displaymath}\langle E_{PK(B)}(A, N_A),  E_{PK(A)}(N_A, N_B)\rangle.\end{displaymath}

The predicate \texttt{InitialKnowledge} states that Oscar may not draw any composite values.  Rather, he must learn them from the protocol rules or the primitives.  The predicate \texttt{FinalKnowledge} specifies  that Oscar's maximal knowledge contains a value $v$ if and only if Oscar draws $v$ or he learns it from a knowledge flow originating in his initial knowledge.

\subsection{Modelling Cryptographic Primitives}

The Needham-Schroeder protocol requires the use of cryptographic primitives to encrypt/decrypt messages and generate nonces.  Our encoding of the knowledge flows and values associated with these primitives is shown in Model Excerpt \ref{primitives}.  Note that we do not explicitly model primitives as principals.  Instead, we define the pools of values drawn by the primitives as signatures and encode their input/output behavior as predicates.  For example, the initial knowledge of $Encryptor$ is given by the set $Ciphertext$, and $Encryptor$'s operation is encoded in the predicates \texttt{Encryptor} and \texttt{Decryptor}.

\newcounter{M2count}
\begin{model}[!h]
\centering
\begin{list}{\scriptsize{\arabic{M2count}}}{\usecounter{M2count} \setlength{\itemsep}{-4pt} 
\setlength{\leftmargin}{13pt} \setlength{\rightmargin}{-6pt}  } 
\tt \frenchspacing
\addtocounter{M2count}{\value{M1count}}
\item module kf/primitives/encryption 
\item open kf/basicdeclarations
\item
\item {\bf sig} Ciphertext {\bf extends} CompositeValue \{
\item \ plaintext: {\bf some} Value, \label{plaintext}
\item \ key: Value \} \label{key}
\item
\item {\bf pred} PublicKeyCryptography() \{ \label{PublicKeyCryptography}
\item \ Ciphertext.key {\bf in} Principal.owns \& AtomicValue \}
\item
\item {\bf pred} Encryptor(x: {\bf set} Value, v : Value) \{
\item \ v {\bf in} Ciphertext \&\& x = v.key + v.plaintext \}
\item 
\item {\bf pred} Decryptor(x: {\bf set} Value, v : Value) \{  
\item \ {\bf some} c : plaintext.v | x = (c.key + c) \&\& 
\item \ \ (PublicKeyCryptography() => \label{sk-constraint}
\item \ \ c.key {\bf in} Oscar.owns) \}
\item 
\item {\bf pred} PerfectCryptography() \{
\item \ ({\bf all} {\bf disj} c1,c2: Ciphertext | c1.plaintext != 
\item \ \ c2.plaintext || c1.key != c2.key)
\item \ ({\bf all} c : Ciphertext | c != c.key \&\& 
\item \ \ c != c.plaintext) \}
\item[$\vdots$]
\item module kf/primitives/nonces
\item open kf/basicdeclarations
\item
\item {\bf sig} Nonce {\bf extends} CompositeValue \{
\item \ seed : Value,
\item \ id : Value \}
\item 
\item {\bf pred} NonceGenerator(x: {\bf set} Value, v : Value) \{
\item \ v {\bf in} Nonce \&\& v.id {\bf in} Oscar.owns \&\& x = v.seed \}
\end{list}

\caption{Cryptographic Values and Primitives}
\label{primitives}
\end{model}

A $Ciphertext$ represents an encryption of a non-empty $plaintext$ (line \ref{plaintext}) with a given $key$ (line \ref{key}).  The predicate \texttt{Encryptor} formalizes the encryption knowledge flow from Oscar's perspective.  It states that, in order to learn the cipher $v$ from the Encryptor, Oscar must provide the input $x$ consisting of the plaintext and the key associated with $v$.  Similarly, the predicate \texttt{Decryptor} stipulates that Oscar can learn the plaintext $v$ after he presents the input $x$ consisting of an encryption of $v$ and the corresponding decryption key.  

Note that this model of ciphers accommodates both public and symmetric key encryption.  Symmetric key encryption is the default; invoking the predicate \texttt{PublicKeyCryptography} switches on public key encryption.  Any atomic value owned by a principal can serve as his public/private key pair.  The public portion of any principal's key is accessible to Oscar through the $draws$ relation.  The decryption constraint on line \ref{sk-constraint} ensures that Oscar can decrypt a message only if he $owns$ the value representing the public/private key pair.   

Nonces are encoded as composites with two fields, $seed$ and $id$.  The field $id$ stores the identity of the principal to whom the nonce was issued.  The predicate \texttt{NonceGenerator} says that, from Oscar's point of view, the generator will issue a nonce labeled with Oscar's identifier when presented with the input seed $x$.     

\subsection{Modelling Protocol Rules}

The models presented so far are a part of a generic Alloy framework developed for analyzing knowledge flows.  This section describes the values and rules specific to the Needham-Schroeder protocol.

\begin{model}[!h]
\centering
\begin{list}{\scriptsize{\arabic{M1count}}}{\usecounter{M1count} \setlength{\itemsep}{-4pt} 
\setlength{\leftmargin}{13pt} \setlength{\rightmargin}{-6pt}  } 
\tt \frenchspacing
\addtocounter{M1count}{\value{M2count}}
\item module kf/needham\_schroeder
\item open kf/basicdeclarations
\item open kf/primitives/encryption
\item open kf/primitives/nonces
\item 
\item {\bf sig} Identity {\bf extends} AtomicValue \{\}\label{identity}
\item
\item {\bf pred} IdentitiesAreKeys() \{
\item \ {\bf all} p : Principal | {\bf some} p.owns \& Identity \&\& \label{unique-ids} 
\item \ Ciphertext.key {\bf in} Identity \} \label{id-keys}
\item 
\item {\bf pred} PrimitiveRules(x : {\bf set} Value, v : Value) \{
\item \ Encryptor(x,v) || Decryptor(x,v) || 
\item \ NonceGenerator(x,v) \}
\item
\item {\bf pred} ProtocolRules(x : {\bf set} Value, v : Value) \{ \label{ns-rules}
\item	\ v {\bf in} Ciphertext \&\& \{
\item \ \  (x : {\bf some} Oscar.draws \&\& 
\item \ \ \ {\bf let} text = v.plaintext, n = text \& Nonce | 
\item \ \ \ \#text = 2 \&\& {\bf one} n \&\& n.seed {\bf in} AtomicValue \&\& 
\item \ \ \ n.id = text \& Identity) ||
\item \ \ (x : {\bf one} Ciphertext \&\& ({\bf some} n : seed.x | 
\item \ \ \ \#x.plaintext = 2 \&\& v.key {\bf in} x.plaintext \&\&
\item \ \ \ n.id = x.key \&\& 
\item \ \ \ v.plaintext = (x.plaintext - v.key) + n)) ||
\item \ \ (x : {\bf one} Ciphertext \&\& 
\item \ \ \ ({\bf some} n : id.(x.key) \& Nonce | 
\item \ \ \ \#x.plaintext = 2 \&\& n {\bf in} x.plaintext \&\&
\item \ \ \ v.plaintext = x.plaintext - n)) \}\}
\item
\item {\bf pred} ApplyRules() \{
\item \ {\bf all} v : Value | {\bf let} x = Oscar.learns.v |
\item \ \ {\bf some} x <=> PrimitiveRules(x, v) || 
\item \ \ \ ProtocolRules(x, v) \}
\end{list}

\caption{Needham-Schroeder Protocol}
\label{protocol-rules}
\end{model}

Principals' identifiers are modelled as atomic values contained in the set $Identity$ (Model Excerpt \ref{protocol-rules}, line \ref{identity}).  Each principal $owns$ an $Identity$ (\ref{unique-ids}), which also doubles as its owners' public/private key pairs (\ref{id-keys}).

The \texttt{ProtocolRules} predicate (line \ref{ns-rules}) embeds the knowledge flow rules given by equations \ref{ns1}-\ref{ns3} into first-order logic.  The predicate \texttt{ApplyRules} states that the $learns$ relation may map the set of values $x$ to the value $v$ if and only if the protocol or primitive rules define a knowledge flow from $x$ to $v$.

\subsection{Checking Security}

The predicate \texttt{SecurityAssumptions} in Model Excerpt \ref{security-theorem} models our assumptions about the properties of cryptographic primitives and principals.  We assume perfect public key cryptography (line \ref{perfect-pk-crypto}) and the use of identifiers as public/private key pairs (line \ref{ids-as-keys}).  

The security property that the protocol should satisfy is given by the predicate \texttt{SecurityTheorem}.  It states that Oscar's maximal knowledge never contains two nonces, $nA$ and $nB$, such that $nB$ is generated by Bob in response to a protocol initialization message sent by Alice (a cipher containing Alice's identity and one of her nonces).  The assertion \texttt{Security} stitches the model together to stipulate that the security property should hold if Oscar obtains his maximal knowledge by applying the knowledge flow rules to the values he draws.

\begin{model}[!h]
\centering
\begin{list}{\scriptsize{\arabic{M2count}}}{\usecounter{M2count} \setlength{\itemsep}{-4pt} 
\setlength{\leftmargin}{13pt} \setlength{\rightmargin}{-6pt}  } 
\tt \frenchspacing
\addtocounter{M2count}{\value{M1count}}
\item {\bf pred} SecurityAssumptions() \{
\item \ PerfectCryptography() \&\& PublicKeyCryptography()\label{perfect-pk-crypto}
\item \ IdentitiesAreKeys() \} \label{ids-as-keys}
\item
\item {\bf pred} SecurityTheorem() \{
\item \ {\bf no} {\bf disj} Alice, Bob : HonestUser, 
\item \ nA, nB : Oscar.knows \& Nonce | 
\item \ \ nA.id {\bf in} Alice.owns \&\& nB.id {\bf in} Bob.owns \&\& 
\item \ \ ({\bf some} c : Ciphertext | nB.seed = c \&\& 
\item \ \ \ c.key = nB.id \&\& 
\item \ \ \ c.plaintext = nA.id + nA) \}
\item
\item {\bf assert} Security \{
\item \ InitialKnowledge() \&\& FinalKnowledge() \&\& 
\item \ SecurityAssumptions() \&\& ApplyRules() => 
\item \ SecurityTheorem() \}
\end{list}

\caption{Security Assumptions and Theorem}
\label{security-theorem}
\end{model}

\begin{figure*}[!h]
\centering
\includegraphics[scale=.5]{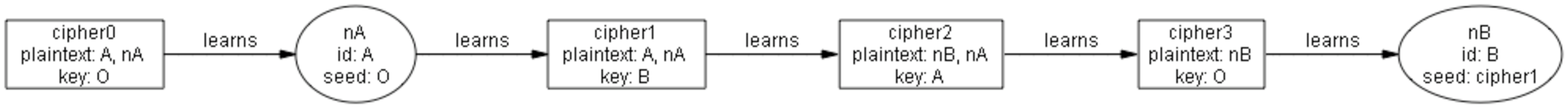}
\caption{Parallel Session Attack on the Needham-Schroeder Protocol}
\label{lowe-attack}
\end{figure*}

The Alloy Analyzer generates a counterexample to the \texttt{Security} assertion (Figure \ref{lowe-attack}) that is a knowledge flow representation of the parallel session attack discovered by Lowe \cite{Lowe96}.  Alice uses $cipher0$ to initiate the protocol with Oscar, who extracts $nA$ and forwards it to Bob in $cipher1$.  Thinking that he is authenticating with Alice, Bob responds with $cipher2$ which Oscar simply forwards to Alice.  She completes the session with Oscar by sending him $nB$, which she believes is his nonce, in $cipher3$.  Oscar now knows both $nA$ and $nB$, contrary to our claim.

\section{Knowledge Flow Analysis} 

Knowledge flow analysis is based on a simple mathematical foundation.  This section formalizes the ideas outlined in the discussion of knowledge flow basics.  We describe how \textit{communication rules} direct knowledge flows (\ref{kf-basic-defs}), show that our treatment of primitives ensures a fixed pool of values (\ref{initial-knowledge}), formulate the analysis problem in terms of Oscar's knowledge flows (\ref{adversary}), and present a small-model theorem which makes our analysis complete for a bounded number of parallel protocol sessions (\ref{kf-completeness}).  

\subsection{Communicating Knowledge}\label{kf-basic-defs}

We denote the sets of all {\em principals} and {\em values} by $P$ and $V$.  A subset of $P\times V$ is a {\em state of knowledge} drawn from $K=2^{P\times V}$, the set of all possible states of knowledge.  For a given state of knowledge $k\in K$, we say that ``$p$ knows $v$'' if $(p,v)\in k$.  
\begin{definition} A tuple $(R,k_0)$ is a {\em knowledge flow} for $(P,V)$ directed by the {\em communication rules} $R\subseteq P\times V\times P\times K$ and originating from the state $k_0\in K$.
\end{definition}
A communication rule describes the conditions under which one principal may gain knowledge from another.  For example, the rule $(e, E_{PK(p_b)}(v), p_a, \{(p_a, PK(p_b)), (p_a, v)\})$ states that the encryptor $e$ will tell the cipher $E_{PK(b)}(v))$ to the principal $p_a$ if $p_a$ knows $p_b$'s public key and the plaintext $v$.  

Note that our definition of a communication rule limits the class of protocols expressible in the knowledge flow framework.  In particular, our rules cannot be used to specify conditions under which information is {\em withheld} from a principal, such as ``$a$ will {\em not} tell $v$ to $b$ if $b$ knows $x$''.  Although many practical protocols do not require this form of expressiveness, withholding of knowledge is an essential concept in systems that use certificates:  revoking a certificate requires withholding of information.  We are working on reformulating the certificate revocation problem using valid and invalid certificate sets, which should allow us to circumvent this limitation.


Given a set of communication rules $R$, we say that $k'\in K$ is reachable from $k\in K$ via $R$ if $k'$ is the result of applying all rules in $R$ to $k$ at most once; i.e. $k'=f_R(k)$ where 
\begin{definition} $f_R: K \longrightarrow K$ such that
$$f_R(k)=k\cup \left\{ (p_a,v) : \begin{array}{l}
(p_b,v) \in k, k_a \subseteq k, \mbox{ and } \\
(p_b,v,p_a,k_a) \in R,\\
\mbox{for some } p_b\in P \mbox{ and } k_a\in K \end{array} \right\}.$$ \label{fR-def}
\end{definition}

A state of knowledge $k_n$ is reachable in the context of a knowledge flow $(R,k_0)$ if $k_n=f_R^n(k_0)$. The {\em maximal state of knowledge} $f^*_R(k_0)$ is the limit of $k_n=f_R^n(k_0)$ as $n\rightarrow \infty$.  A state of knowledge $f^*_{R_\kappa}(\kappa)$  is {\em valid} for a knowledge flow $(R,k_0)$ if $R_\kappa \subseteq R$ and $\kappa\subseteq k_0$.  Since $f_R(k_0)$ is monotonically increasing\footnote{It is evident from Definition \ref{fR-def} that self-rules such as $r=(p,v,p,k_p)\in R$ do not affect the flow of knowledge:  $f_R(k)=f_{R-\{r\}}(k)$. We therefore assume that $R$ does not contain any self-rules.} in $R$ and $k_0$, any valid state of knowledge is a subset of the maximal state of knowledge. Hence, the maximal state of knowledge is also the smallest fixed point of $f_R$ which subsumes $k_0$. 

\subsection{Initial Knowledge}\label{initial-knowledge}

For each value $v$, $Source(v)=\{p: (p,v)\in k_0\}$ defines the set of principals who draw $v$.  In the knowledge flow framwork, a principal $p$ outside of $Source(v)$ can learn $v$ only by communicating with principals who know $v$.  We therefore treat cryptographic primitives, and other computationally feasible algorithms, as principals. For example, suppose that, in practice, $p$ can compute $v$ by applying the algorithm ${\cal A}$ to inputs $i_1, i_2, \ldots, i_n$.  We model ${\cal A}$ by adding the principal $A$ to $P$, the tuple $(A,v)$ to $k_0$, and the rule $(A,v,p,\{(p,i_1),(p,i_2), \ldots (p, i_n)\})$ to $R$.

Our treatment of primitives ensures that $Knowledge(k_0)=\{v: (p,v)\in k_0 \mbox{ for some } p\in P\}$
consists of {\em all} learnable values. Hence, $V$ is the same in the initial and the maximal state of knowledge,
\begin{equation} Knowledge(k_0) = Knowledge(f^*_R(k_0)), \label{v0} \end{equation}
which implies that we can safely restrict our analysis to the subset of $R$ applicable to $k_0$.  Formally,
$$(\ref{v0}) \implies f_R(k_0)=f_{R(k_0)}(k_0) \mbox{ and } f^*_R(k_0)=f^*_{R(k_0)}(k_0),$$
where $R(k_0) = \left\{ (p_b,x,p_a,k_a)\in R : \begin{array}{l}\{x\}\cup \{v:(p_a,v)\in k_a\} \\ \subseteq Knowledge(k_0) \end{array}\right\}.$

\subsection{Adversaries' Knowledge}\label{adversary}

Let $O\subseteq P$ be a group of collaborating adversaries. We collapse $O$ into a single principal $o$ using the following merging function:
$$\begin{array}{rcl}
Merge(p) &=& \left\{\begin{array}{l}o \mbox{ if } p\in O, \\ p \mbox{ if } p\not\in O
\end{array}\right.\\
Merge(k) &=& \{(Merge(p),v) : (p,v) \in k \}\\
Merge(r) &=& (Merge(p_b),v,Merge(p_a),Merge(k_a))\\
&&\mbox{where } r = (p_b,v,p_a,k_a) \in R
\end{array}$$
The merging of adversaries does not rule out any attacks because $Merge(f^*_R(k_0))\subseteq f^*_{Merge(R)}(Merge(k_0))$.  We subsequently assume that $Merge$ is implied and use $P$, $R$, and $k_0$ to refer to $Merge(P)$, $Merge(R)$, and $Merge(k_0)$.

Security properties of protocols are expressed as predicates on the values known to Oscar in the maximal state of knowledge.  We therefore restrict our analysis of knowledge flows to finding all the values in the projection of $f^*_{R(k_0)}(k_0)$ on Oscar.  Specifically, we introduce the projection function $g_{R,k_0}$ and show that its smallest fixed point is the image of Oscar under $f^*_{R(k_0)}(k_0)$.
\begin{definition} \label{defarrow} Let $X \rightarrow x$ denote the existence of a rule $(p,x,o,k_\sigma)\in R(k_0)$ for some $p\in P-\{o\}$ and $k_\sigma\in K$ with $X=\{v :(o,v)\in k_\sigma\}$.  We define  $g_{R,k}: 2^V \longrightarrow 2^V$ as
$$g_{R,k_0}(X)=X\cup \left\{ x : X_\sigma \rightarrow x 
\mbox{ for some } X_\sigma\subseteq X \right\}.$$
 The set of values reachable from $X$ is given by $g^*_{R,k_0}(X)$, which is the limit of $g_{R,k_0}^n(X)$ as $n\rightarrow \infty$.\label{gR-def}
\end{definition}

Since $f_R(k_0)$ is monotonically increasing in $R$ and $k_0$, Oscar's pool of values under $f^*_R(k_0)$ is maximized if (a) Oscar tells everything he knows to the honest principals and (b) the honest principals tell everything they know to each other.  Therefore, $(P-\{o\})\times Knowledge(k_0)$ should be included in the maximal state of knowledge.  This is equivalent to assuming that each honest principal draws $Knowledge(k_0)$ because  $k\subseteq f^*_R(k_0)$ implies that $f^*_R(k_0)=f^*_R(k_0\cup k)$. 

\begin{lemma} \label{arrow}
Let $[(P-\{o\})\times V_0]\subseteq k_0$ with $V_0=Knowledge(k_0)$ and let $k_n=f^n_R(k_0)$. Then there exists a unique set $X_n\subseteq V$ such that 
\begin{equation}
k_n=[(P-\{o\})\times V_0] \cup [\{o\}\times X_n].\label{xn}
\end{equation}
The set $X_n$ has the property that $X_n=g^n_{R,k_0}(X_0)$.
\end{lemma}

\noindent
{\em Proof.}

We use induction on $n$.  For $n=0$, $X_n=X_0=g^n_{R,k_0}(X_0)$. Since $(P-\{o\})\times V_0\subseteq k_0$ and $V_0=Knowledge(k_0)$, there exists a unique $X_0$ such that $k_0$ satisfies (\ref{xn}).  
 
Let $X_n=g^n_R(X_0)$ be a unique solution to (\ref{xn}) and \\$Knowledge(k_n)=V_0$ (our induction hypothesis).  We know that $k_{n+1}=f_R(k_n)$ and, therefore, $Knowledge(k_{n+1})=Knowledge(k_n)=V_0$. Together with $[(P-\{o\})\times V_0] \subseteq k_n\subseteq k_{n+1}$, this implies the existence a unique $X_{n+1}$ for which $k_{n+1}$ satisfies (\ref{xn}).  We now need to prove that $X_{n+1}=g^{n+1}_{R,k_0}(X_0)$.

Definition (\ref{xn}) lets us infer that  $x\in X_{n+1} \Longleftrightarrow (o,x)\in k_{n+1}=f_R(k_n)$.  According to Definition (\ref{fR-def}), $(o,x)\in f_R(k_n)$ if and only if i) $(o,x)\in k_n$, which is, by (\ref{xn}),  equivalent to $x\in X_n$, or ii) there exists a $p\in P$ and $k_\sigma\in K$ such that $(p,x)\in k_n$, $k_\sigma\subseteq k_n$, and $(p,x,o,k_\sigma)\in R$. Since there are no self-rules $(o,v,o,k_\sigma)\in R$, we know that $p\in P-\{o\}$.  This, together with $x\in X_{n+1}\subseteq V_0$, implies that $(p,x)\in [(P-\{o\})\times V_0] \subseteq k_n$. Given $[(P-\{o\})\times V_0] \subseteq k_n$ and $V_0=Knowledge(k_n)$, the condition $k_\sigma\subseteq k_n$ is equivalent to 
$$\begin{array}{l} X_\sigma=\{v: (o,v)\in k_\sigma\} \subseteq \{v: (o,v)\in k_n\} =X_n \mbox{ and}\\ \{v:(o,v)\in k_\sigma\}\subseteq V_0.\end{array}$$ 
Since $x\in X_{n+1}\subseteq V_0$, $\{v:(o,v)\in k_\sigma\}\subseteq V_0$ gives us $(p,x,o,k_\sigma)\in R(k_0)$. 
Therefore, case ii) holds if and only if there exists a set $X_\sigma\subseteq X_n$  such that $X_\sigma\rightarrow x$. By Definition (\ref{gR-def}), case i) or case ii) holds if and only if $x\in g_{R,k_0}(X_n)$.  Hence, $X_{n+1}=g_{R,k_0}(X_n)$ and the lemma follows by induction on $n$.

\hfill $\Box$

\subsection{Detecting Intruders}\label{kf-completeness}
Let $m$ be the total number of values used in a single protocol session, including the subterms of each composite value. Suppose that Oscar can use only the primitives which  {\em compose} or {\em decompose} inputs and for which the composition rules have no collisions (e.g encryptor/decryptor). Then, the theory in \cite{RT2001} implies the following: if there exists an attack in which Oscar uses $w$ parallel protocol sessions, then such an attack need not involve more than $w\cdot m$ values.
 From (\ref{v0}) we infer that this corresponds to a valid state of knowledge $f_R^*(k_\sigma)$ derived from the set $k_\sigma\subseteq k_0$ of cardinality $|Knowledge(k_\sigma)|\leq wm$. By Lemma \ref{arrow}, we can conclusively {\em decide} whether there is an attack which uses $w$ parallel protocol sessions by computing
\begin{equation} \left\{v\in g^*_{R,k_\sigma}(X_\sigma) : 
\begin{array}{l} \mbox{for }  
[\{o\}\times X_\sigma] \subseteq k_\sigma \subseteq k_0\\
\mbox{with } |Knowledge(k_\sigma)|\leq wm  
\end{array} 
\right\}. \label{comp}
\end{equation}

\section{Evaluation}

We have applied the theory developed in the previous section to check the security of the original \cite{NS78} and modified \cite{Lowe96} Needham-Schroeder Public Key Protocol, the Otway-Rees Mutual Authentication Protocol \cite{Otway-Rees87}, and the bootstrapping and renewal protocols based on Controlled Physical Random Functions (CPUFs) \cite{Gassend03PhysicalRandomFunctions, GassendETAL02ControlledPhysicalRandomFunctions}.  

The knowledge flows of the protocols were embedded into Alloy using the pattern presented in section \ref{example}.  The pattern is embodied in a general Alloy framework for knowledge flow analysis which includes definitions of basic concepts (Model Excerpt \ref{basicdeclarations}), a library of primitives, and a model outline for specifying protocol rules and security theorems.  For example, Model Excerpt \ref{primitives} shows portions of Alloy modules that encode generic encryption/decryption and nonce generator primitives, and Model Excerpts \ref{protocol-rules} and \ref{security-theorem} comprise an instantiation of the modelling outline for the Needham-Schroeder protocol. 

We have found that the Alloy framework and its associated tool support make the process of  knowledge flow modelling fast, simple, and accurate.  Our analysis is sound and, since most cryptographic primitives used in practice are composing/decomposing, we can make it complete for a bounded number of parallel sessions by applying the results from section \ref{kf-completeness}.  In the case of the modified Needham-Schroeder protocol, for example, we have proved that it is secure against all attacks that use two parallel sessions.  The analysis of the Otway-Rees protocol (\ref{otway-reese-protocol}) produced the type flaw attack described in \cite{BurrowsETAL90LogicAuthentication}.  We found the CUPFs protocols (\ref{cpuf-protocol}) to be secure for a single protocol session and, therefore, for an unlimited number of sessions.   

The main limitation of our approach is that it is not fully general.  As pointed out in section \ref{kf-basic-defs}, protocols that {\em withhold} information under certain conditions cannot be formulated as knowledge flows.  However, this limitation does not significantly detract from practical usefulness of knowledge flow analysis:  as far as we know, few practical protocols contain information-withholding rules.     

\section{Related Work}

The first formalisms
designed for reasoning about cryptographic protocols are belief logics such as
BAN logic \cite{BurrowsETAL90LogicAuthentication},
used by the Convince tool 
\cite{LichotaETAL96VerifyingCryptographicProtocolsElectronicCommerce} 
with the HOL theorem prover \cite{GordonMelham93HOL}, and its generalizations (GNY \cite{GNY90}, AT \cite{AT91}, and  SVO logic \cite{SO94}
which the C3PO tool \cite{Dekker00C3PO} employs with the Isabelle theorem prover
\cite{NipkowETAL02IsabelleHOLTutorial}).
Belief logics are difficult to use since the logical form of a protocol does not correspond to the protocol itself in an obvious way.  Almost indistinguishable formulations of the same problem lead to different results. It is also hard to know if a formulation is over constrained or if any important assumptions are missing. BAN logic and its derivatives cannot deal with security flaws resulting from interleaving of protocol steps \cite{BM93} and cannot express any properties of protocols other than authentication \cite{MB93}.
To overcome these limitations, the knowledge flow formalism has, like other approaches \cite{Lowe97,MMS97,CJM00,SongETAL01AthenaNovelApproachEfficientAutomatic,Meadows94NRLProtocolAnalyzer}, a concrete operational model of protocol execution.  Our model also includes a description of how the honest participants in the protocol behave  and a description of how an adversary can interfere with the execution of the protocol. 

Specialized model checkers such as Casper \cite{Lowe97}, Mur$\phi$ \cite{MMS97},  Brutus \cite{CJM00}, TAPS \cite{Cohen04TAPS}, and ProVerif \cite{AB} have been successfully used to analyze security protocols.  Like knowledge flow analysis in Alloy, these tools are based on state space exploration which leads to an exponential complexity. Athena \cite{SongETAL01AthenaNovelApproachEfficientAutomatic} is based on a modification of the strand space model \cite{FabregaETAL98StrandSpaces}. Even though it reduces the state space explosion problem, it remains exponential.  Multiset rewriting \cite{DurginETAL04MultisetRewritingSecurity} in combination with tree automata is used in Timbuk \cite{timbuk}. The relation between multiset rewriting and strand spaces is analyzed in \cite{MR-SP}. The relation between multiset rewriting and process algebras \cite{pi,spi} is analyzed in \cite{MR-PA}. 

Proof building tools such as NRL, based on Prolog \cite{Meadows94NRLProtocolAnalyzer}, have also been helpful for analyzing security protocols. However, they are not fully automatic and often require extensive user intervention. Model checkers lead to completely automated tools which generate counterexamples if a protocol is flawed. For theorem-proving-based approaches, counterexamples are hard to produce. 

For completeness, we note that if the initial knowledge of the intruder consists of a finite number of explicit (non-parameterized, non-symbolic) values, then a polynomial time intruder detection algorithm can be shown to exist using a generalization of the proof normalization arguments
\cite{McAllester93AutomaticRecognitionTractabilityInferenceRelations,
BasinGanzinger01AutomatedComplexityAnalysisBasedOrderedResolution,
GivanMcallester02PolynomialInference}, 
which were employed in \cite{BodeiETAL02FlowDolevYao,
NielsonETAL01CryptographicAnalysisCubicTime} 
and have been implemented in the framework 
\cite{NielsonETAL04SuccinctSolverSuite}. However, in practice,  the initial knowledge of an intruder is unbounded and represented by a finite number of parameterized sets, each having an infinite number of elements.

The key advantage of the knowledge flow approach over other formalisms is its simplicity and flexibility. It is simple in the sense that the underlying mathematics is straightforward and elementary; it does not require any specialized background (in logic). It is flexible in the sense that the same library of cryptographic primitives can be used to model different protocols and that the security of a complex scheme involving multiple protocols can be verified. Knowledge Flow Analysys allows modeling of confidentiality and authenticity via a wide range of primitives such as pairing, union, hashing, symmetric key encryption, public key encryption, MACs and digital signatures.

Our formalism derives its simplicity from being just sufficiently expressive to enable modelling of practical cryptographic protocols.  In particular, existentials \cite{DurginETAL04MultisetRewritingSecurity} cannot be encoded as knowledge flows; existentials are implicitly modeled in Oscar's initial knowledge. As mentioned in Section (\ref{kf-basic-defs}), NP-hardness proofs which use (existential) Horn clause reduction \cite{DurginETAL04MultisetRewritingSecurity} or SAT3 reduction  \cite{RT2001} are not applicable to Knowledge Flow Analysis.

\section{Conclusion}

This paper introduces a new method for formalizing and checking security protocols.  Our approach enables natural encoding of protocol rules, simple treatment of primitives, direct embedding into first order logic, and sound analysis that is also complete for many practical protocols.   

We have developed a general framework for analyzing knowledge flows using the Alloy Analyzer.  The framework has been used to generate easily understandable knowledge flow representations of  parallel session and type flaw attacks on the Needham-Schroeder and Otway-Rees protocols.  We have also instantiated it with the rules for CPUFs key management protocols and verified the protocols' correctness for an unlimited number of parallel sessions.

We believe that knowledge flow analysis may be polynomial-time decidable for some protocols.  Future work will involve identifying the class of protocols whose knowledge flows are analyzable in polynomial time and developing a specialized tool for checking them.

\section*{Acknowledgments}
We would like to thank Viktor Kuncak, Ishan Sachdev, and Ilya Shlyakhter for their contributions to and comments on earlier versions of this work.

\bibliographystyle{abbrv}

\setcounter{section}{0}
\renewcommand{\thesection}{Appendix \Alph{section}}
\section{The Otway-Rees Protocol}\label{otway-reese-protocol}

\begin{list}{\scriptsize{\arabic{M1count}}}{\usecounter{M1count} \setlength{\itemsep}{-4pt} 
\setlength{\leftmargin}{13pt} \setlength{\rightmargin}{-6pt}  } 
\tt \frenchspacing
\item {\bf module} kf/otwayreese
\item {\bf open} kf/basicdeclarations
\item {\bf open} kf/primitives/encryption
\item {\bf open} kf/primitives/nonces
\item
\item {\bf sig} Message  {\bf extends} CompositeValue \{
\item \ contents:  {\bf some} Value \}
\item
\item  {\bf sig} Identity  {\bf extends} AtomicValue \{\}
\item
\item {\bf pred} PrimitiveRules(x : {\bf set} Value, v : Value) \{
\item \ Encryptor(x,v) || Decryptor(x,v) || NonceGenerator(x,v) ||
\item \ (x : Message \&\& v {\bf in} x.contents) \}
\item
\item {\bf pred} idCipher(cipher: Value) \{
\item \ cipher : Ciphertext	\&\&		
\item \ {\bf some} cipher.key.id \& cipher.plaintext \&\&
\item \ cipher.plaintext {\bf in} Identity \&\&
\item \ {\bf one} cipher.plaintext - cipher.key.id \}
\item 
\item {\bf pred} keyCipher(cipher: Value) \{
\item \ cipher : Ciphertext \&\&
\item \ {\bf some} cipher.key.id \}
\item
\item {\bf pred} message1(m: Value) \{
\item \ m : Message \&\&
\item \ {\bf let} cipher = m.contents \& Ciphertext | \{
\item \ \ idCipher(cipher) \&\&
\item \ \ m.contents = cipher + cipher.plaintext \}\}
\item 
\item {\bf pred} message2(m: Value) \{
\item \ m : Message  \&\&
\item \ {\bf some} cipher1 : Ciphertext | 
\item \ \ {\bf let} cipher2 = m.contents \& Ciphertext - cipher1 | \{
\item \ \ \  idCipher(cipher1) \&\&
\item \ \ \  idCipher(cipher2) \&\&
\item \ \ \  cipher1.plaintext = cipher2.plaintext \&\&
\item \ \ \  m.contents = cipher1 + cipher2 + cipher1.plaintext \}\}
\item 
\item {\bf pred} message3(m: Value) \{
\item \ m : Message \&\&
\item \ {\bf some} cipher1 : Ciphertext | 
\item \ \ {\bf let} cipher2 = m.contents \& Ciphertext - cipher1 | \{
\item \ \ \  keyCipher(cipher1) \&\&
\item \ \ \  keyCipher(cipher2) \&\&
\item \ \ \  cipher1.plaintext = cipher2.plaintext \&\&
\item \ \ \  m.contents = cipher1 + cipher2 \}\}
\item 
\item {\bf pred} message4(m: Value) \{
\item \ m : Message \&\&
\item \ keyCipher(m.contents) \}
\item 
\item {\bf pred} ProtocolRules(x : {\bf set} Value, v : Value) \{
\item \ (x : {\bf some} Oscar.draws \&\& message1(v)) ||
\item \ (message1(x) \&\& message2(v) \&\& x.contents {\bf in} v.contents) ||
\item \ (message2(x) \&\& message3(v) \&\& x.contents.key = v.contents.key) ||
\item \ (message3(x) \&\& message4(v) \&\& v.contents {\bf in} x.contents) \}
\item 
\item {\bf pred} ApplyRules() \{
\item \ {\bf all} v : Value | {\bf let} x = Oscar.learns.v |
\item \ \ {\bf some} x <=> PrimitiveRules(x, v) || ProtocolRules(x, v) \}
\item
\item  {\bf pred} SecurityAssumptions()\{
\item \ PerfectCryptography() \&\&
\item \ {\bf all} p : Principal | {\bf some} p.owns \& Identity \}
\item
\item  {\bf pred} SecurityTheorem() \{
\item \ {\bf no} oldResp, newResp : PUFResponse,  
\item \ renew : param.(oldResp.isRespTo) \& HonestUser,
\item \ cipher : Ciphertext  | 
\item \ \ {\bf let} oldChall = oldResp.isRespTo, newChall = newResp.isRespTo | 
\item \ \ \ oldChall.isHashOf : {\bf some} (AtomicValue - Oscar.draws) \&\& 
\item \ \ \ cipher.key.isHashOf = oldResp + renew.hash \&\&
\item \ \ \ cipher.plaintext = newResp \&\&
\item \ \ \ newChall.isHashOf = oldChall + renew.hash \&\&
\item \ \ \ newResp {\bf in} Oscar.knows \}
\item
\item {\bf assert} Security \{
\item \ InitialKnowledge() \&\& FinalKnowledge() \&\& 
\item \ SecurityAssumptions() \&\& ApplyRules() => SecurityTheorem() \}
\item
\item {\bf pred} SecurityTheorem() \{
\item \ {\bf no} m1, m2, m3, m4: Oscar.knows \& Message, 
\item \ A, B: HonestUser.owns \& Identity | \{
\item \ \ message1(m1) \&\& message2(m2) \&\& 
\item \ \ message3(m3) \&\& message4(m4) \&\&
\item \ \ m1.contents.key.id = A \&\&
\item \ \ m2.contents.key.id = A + B \&\&
\item \ \ m3.contents.key.id = A + B \&\&
\item \ \ m4.contents.key.id = A \&\&
\item \ \ m4.contents.plaintext {\bf in} 
\item \ \ \ (HonestUser.draws - Oscar.draws) \& AtomicValue \&\&
\item \ \ m4.contents.plaintext {\bf in} Oscar.knows \}\}
\item
\item {\bf assert} Security \{
\item \ InitialKnowledge() \&\& FinalKnowledge() \&\& 
\item \ SecurityAssumptions() \&\& ApplyRules() => SecurityTheorem() \}
\end{list}

\section{The CUPF Renewal Protocol}\label{cpuf-protocol}

\begin{list}{\scriptsize{\arabic{M1count}}}{\usecounter{M1count} \setlength{\itemsep}{-4pt} 
\setlength{\leftmargin}{13pt} \setlength{\rightmargin}{-6pt}  } 
\tt \frenchspacing
\item {\bf module} kf/primitives/hashing
\item {\bf open} kf/basicdeclarations
\item
\item {\bf sig} Hash  {\bf extends} CompositeValue \{
\item \ isHashOf:  {\bf some} Value \}
\item
\item {\bf pred} CollisionFreeHashing() \{
\item \ {\bf all disj} h1, h2: Hash | h1.isHashOf != h2.isHashOf \}
\item
\item {\bf pred} Hasher(x : {\bf set} Value, v : Value) \{
\item \ v {\bf in} Hash \&\& x = v.isHashOf \}
\end{list}

\begin{list}{\scriptsize{\arabic{M2count}}}{\usecounter{M2count} \setlength{\itemsep}{-4pt} 
\setlength{\leftmargin}{13pt} \setlength{\rightmargin}{-6pt}  } 
\tt \frenchspacing
\addtocounter{M2count}{\value{M1count}}
\item {\bf module} kf/cpufs
\item {\bf open}  kf/basicdeclarations
\item {\bf open}  kf/primitives/encryption
\item {\bf open}  kf/primitives/hashing
\item 
\item {\bf sig} PUFResponse {\bf extends} CompositeValue \{
\item \ isRespTo: Value \}
\item
\item {\bf pred} UniquePUFResponses() \{
\item \ {\bf all} r: PUFResponse | r.isRespTo !{\bf in} (PUFResponse - r).isRespTo  \}
\item 
\item {\bf sig} RenewProg {\bf in} Principal \{
\item \ param : Value,
\item \ hash : AtomicValue \& owns
\item  \}\{ param + hash  {\bf in} draws + knows \}
\item 
\item  {\bf pred} SecretsNotLeaked() \{
\item \ {\bf no} (RenewProg \& HonestUser).param.isHashOf \& PUFResponse \&\&
\item \ (RenewProg \& HonestUser).param {\bf in} Hash \}
\item 
\item  {\bf pred} GetResponsePrimitive(x : {\bf set} Value, v : Value) \{
\item \ v {\bf in} PUFResponse \&\& 
\item	\ v.isRespTo.isHashOf = Oscar.hash + Oscar.param \&\& 
\item	\ x = v.isRespTo \}
\item 
\item {\bf pred} GetSecretPrimitive(x : {\bf set} Value, v : Value) \{
\item \ v {\bf in} Hash \&\& 
\item \ v.isHashOf = isRespTo.(Oscar.param) + Oscar.hash \&\&
\item \ x = v.isHashOf \} 
\item
\item {\bf pred} PrimitiveRules(x : {\bf set} Value, v : Value) \{
\item \ Encryptor(x,v) || Decryptor(x,v) || 
\item \ GetResponsePrimitive(x,v) || GetSecretPrimitive(x,v) \}
\item
\item {\bf pred} ProtocolRules(x : {\bf set} Value, v : Value) \{
\item \ x : {\bf some} Oscar.draws \&\& \{
\item \ \ (v {\bf in} (RenewProg \& HonestUser).(param + hash)) || 
\item \ \ (v {\bf in} Ciphertext \&\&
\item \ \ \ {\bf some} renew: RenewProg \& HonestUser | 
\item \ \ \ \ {\bf let} renewHash = renew.hash | 
\item \ \ \ \ \ v.key.isHashOf = isRespTo.(renew.param) + renewHash \&\&
\item \ \ \ \ \ v.plaintext.isRespTo.isHashOf = renewHash + renew.param) \}\}
\item
\item {\bf pred} ApplyRules() \{
\item \ {\bf all} v : Value | {\bf let} x = Oscar.learns.v |
\item \ \ {\bf some} x <=> PrimitiveRules(x, v) || ProtocolRules(x, v) \}
\item
\item  {\bf pred} SecurityAssumptions()\{
\item \ UniquePUFResponses() \&\& PerfectCryptography() \&\&
\item \ SingleValueEncryption() \&\& CollisionFreeHashing() \&\&
\item \ SecretsNotLeaked() \}
\item
\item  {\bf pred} SecurityTheorem() \{
\item \ {\bf no disj} oldResp, newResp : PUFResponse,  
\item \ renew : param.(oldResp.isRespTo) \& HonestUser,
\item \ cipher : Ciphertext  | 
\item \ \ {\bf let} oldChall = oldResp.isRespTo, newChall = newResp.isRespTo | 
\item \ \ \ oldChall.isHashOf : {\bf some} (AtomicValue - Oscar.draws) \&\& 
\item \ \ \ cipher.key.isHashOf = oldResp + renew.hash \&\&
\item \ \ \ cipher.plaintext = newResp \&\&
\item \ \ \ newChall.isHashOf = oldChall + renew.hash \&\&
\item \ \ \ newResp {\bf in} Oscar.knows \}
\item
\item {\bf assert} Security \{
\item \ InitialKnowledge() \&\& FinalKnowledge() \&\& 
\item \ SecurityAssumptions() \&\& ApplyRules() => SecurityTheorem() \}
\end{list}

\end{document}